# Routing Around the Queue: An Evaluation of Eco-Routing Protocol on different Traffic Scenarios

Leandros Maglaras[1], George Boustras[2], Azidine Guezzaz[3], Naghmeh Moradpoor[4]

[1]*School of Computer Science and Informatics, De Montfort University, Leicester, UK*

[2]*School of Business Administration and Digital Technologies, European University Cyprus*

[3]*SISAR Team, LaRTiD Laboratory, Technology Higher School Essaouira, Cadi Ayyad University, Morocco*

[4]*School of Computing, Engineering & The Built Environment, Edinburgh Napier University, UK*

***Abstract***—We evaluate ErouVe, a vehicular eco-routing protocol in which roadside units (RSUs) collect per-segment travel-time and $CO_2$ reports from passing vehicles and answer route requests with the lowest-cost alternative, across a range of road architectures and traffic scenarios. We conduct the evaluation with a working implementation of the protocol inside Veins, adapted to drive SUMO's own router rather than a hand-rolled shortest-path search, on a network offering a short, signal-controlled route and a longer, signal-free bypass, under two traffic scenarios. This evaluation reveals clear strengths and weaknesses of the protocol. Under light, uncongested traffic, ErouVe's fleet-wide effect is light. Under traffic dense enough to stress the signalized route's throughput and produce a real, growing queue, the same protocol succeeds a medium fleet-wide $CO_2$ reduction, with diverted vehicles individually saving even more fuels. We use these findings to propose three targeted corrective methods; a knowledge-gated cost-acceptance rule, a diversion-locking rule, and a distance-normalized $CO_2$ cost function, that respectively address a cold-start knowledge asymmetry, decision flapping under noisy crowd-sourced data, and a per-second $CO_2$ metric that rewards idling. Each of these weaknesses are generic pitfalls for any crowd-sourced routing signal and all proposed methods measurably restore protocol effectiveness.

***Index Terms***—vehicular ad hoc networks, eco-routing, V2X communications, SUMO, traffic congestion, $CO_2$ emissions.

## I. INTRODUCTION

Road transport is one of the more stubborn pieces of the urban $CO_2$ budget, and a large share of it is not about the distance travelled but about how badly it is travelled: idling at lights, crawling behind a queue, accelerating out of a stop that a slightly different route would have avoided entirely. Vehicular ad hoc networks (VANETs) made it plausible, over a decade ago, that a vehicle's on-board unit could report exactly this kind of information to nearby infrastructure in real time, and that the infrastructure could hand back something more useful than a static map: a route weighted by what is actually happening on the road right now [1], [7]–[10].

It is worth being precise about who this still matters for. Electric vehicles are growing fast and unevenly: they reached roughly a quarter of new car sales worldwide in 2025, and passed half of new sales in China specifically [5]. New-sales share is not fleet share, however. The same reporting puts the electrified share of the existing global vehicle stock — the cars actually on the road — at only around 5% [6]. In other words, more than nineteen in twenty vehicles currently driving through a city still burn petrol or diesel. An eco-routing layer aimed at combustion-engine emissions is therefore not chasing a shrinking population; for the foreseeable future it addresses most of what is actually on the road.

In this paper we evaluate ErouVe, the eco-routing protocol proposed by Maglaras, Basaras and Katsaros [1], under road architectures and traffic scenarios that its original evaluation, and the robustness follow-up that extends it [2], do not test directly. To conduct this evaluation we built a working, running implementation of the protocol inside the Veins/OMNeT++/SUMO simulation stack, which we use as an evaluation instrument rather than as the paper's contribution in itself. Running the identical protocol over the identical two-route network under light and then deliberately heavy traffic exposes a question neither prior study isolates: how much does the ambient traffic condition itself determine whether eco-routing has anything to offer? Using these scenarios we identify both the strong and weak aspects of ErouVe, and we use them to propose targeted new methods that address the weaknesses we found. Concretely, this paper makes the following contributions:

- We evaluate ErouVe under two traffic regimes on a specific-built two-route network, exposing a congestion dependency in its effectiveness that previous works didn't examine.
- From this evaluation we identify one clear strength (a 3.28% fleet-wide $CO_2$ reduction under saturated traffic) and three concrete weaknesses of the protocol: a cold-start knowledge asymmetry, decision flapping under noisy crowd-sourced reports, and a per-second $CO_2$ metric that rewards idling rather than penalizing it.
- We propose and evaluate three targeted corrective methods, one for each weakness identified, and show that each measurably restores protocol effectiveness.
- We present current limitations of our proposed methods and present future direction for further corrections and analysis.

The remainder of the paper is organized as follows. Section II reviews related work. Section III summarizes the ErouVe protocol. Section IV describes our evaluation testbed, the weaknesses we identify while exercising it, and the corrective methods we propose. Section V details the experimental setup, Section VI presents our results, Section VII discusses their implications, and Section VIII concludes the article and proposes future directions of research.

## II. RELATED WORK

Inter-vehicle communication has been explored for safety, traffic management, and infotainment applications for well over a decade, with eco-routing and emission reduction forming one recurring thread. Santamaria *et al.* combine cooperative awareness with routing to jointly improve safety and reduce $CO_2$ in VANETs [7], while Barba *et al.* couple warning messages and traffic statistics with intelligent traffic-light control to the same end [8]. Tsugawa and Kato situate vehicular eco-communication within the broader Energy ITS agenda [9], and de Souza *et al.* show that an intelligent traffic information system built on inter-vehicle communication can measurably decrease greenhouse-gas emissions in simulation [10]. Congestion estimation itself, the quantity that ultimately determines whether any of these systems have something to correct, has been approached through cooperative, RSU-mediated measurement rather than fixed historical models, as in the distributed congestion-quantification scheme of Milojevic and Rakocevic [11]. Because a crowd-sourced routing signal is only as trustworthy as the vehicles reporting it, misbehavior and intrusion detection for VANETs is a closely related concern, addressed for example by Maglaras' distributed detection system [12].

Our starting point, ErouVe [1], sits inside this landscape: RSUs accumulate per-segment travel-time and $CO_2$ reports from passing vehicles, exchange them with neighboring RSUs, and answer route requests from approaching vehicles with a cost-weighted recommendation. A follow-up study by the same group plus Janicke [2] examines ErouVe's robustness to malicious or tweaked data and proposes a majority-voting variant of the same three-feature comparison, together with plausibility- and consistency-based filtering of vehicle reports. We do not implement that robustness layer here (our vehicles are non-adversarial throughout) but we return to it in Section VIII as a natural next step. What distinguishes the present work from both [1] and [2] is that neither evaluates the protocol as a function of ambient congestion: both hold traffic conditions effectively fixed and vary the presence or absence of an attacker. We instead hold the protocol and the attack surface fixed and vary the traffic regime and road architecture, which is the evaluation gap this paper addresses.

More recent work reinforces that both halves of this gap, congestion-dependent routing and the trustworthiness of the crowd-sourced signal it depends on, remain active. Talaat *et al.* propose a fog-based V2X traffic management architecture that pairs dynamic routing with incentive mechanisms for sustainable driving behavior [13], while Alabdouli *et al.* survey how route-guidance systems integrate with evolving V2X communication technologies [14]. Wang *et al.* use SUMO to evaluate speed-restriction strategies for alleviating congestion in a tunnel environment, underscoring that a road architecture's own bottlenecks, not just its traffic volume, shape how much a routing intervention can achieve [15]; Liang *et al.* similarly use SUMO to coordinate connected-vehicle rerouting around dedicated bus lanes in mixed traffic [16]. On the trust side, Mármol Campos *et al.* and Mansouri *et al.* both propose federated-learning-based misbehavior detection for intelligent transportation systems [17], [18], a more recent counterpart to the plausibility- and consistency-based filtering used in ErouVe's own robustness follow-up [2]. None of these recent studies, however, isolate ambient congestion itself as the variable that determines whether a routing protocol has anything to offer, which is the gap we target.

## III. THE EROUVE PROTOCOL

ErouVe organizes its exchange of information into three phases, run continuously between vehicles and roadside units (RSUs), within the decentralized control-range architecture shown in Fig. 1 and summarized in Fig. 2:

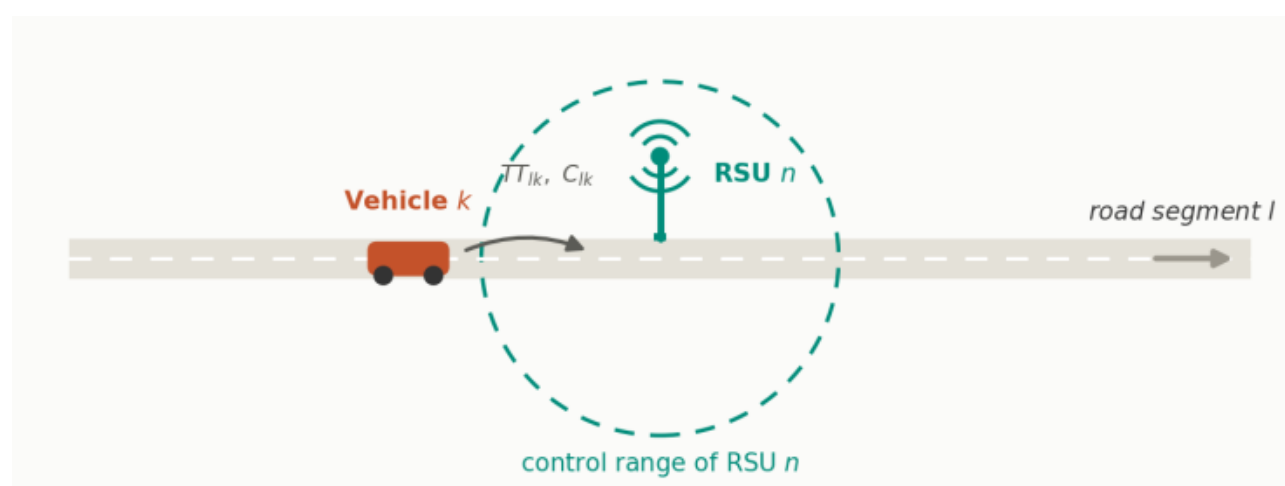

Fig. 1. Decentralized eco-routing architecture: a vehicle inside an RSU's control range reports its segment measurements over DSRC.

**Phase 1: I2V, segment measurement.** When a vehicle finishes traversing a road segment, its serving RSU has it report the travel time and $CO_2$ emitted while crossing that segment. For a vehicle $k$ that has just traversed segment $l$, these two raw quantities are denoted $TT_{lk}$ (its travel time on $l$, in seconds) and $C_{lk}$ (the $CO_2$ it emitted while crossing $l$). The serving RSU aggregates these per-vehicle reports into a running mean for the segment,

$$X_l = (1 / |M(l)|) \ \Sigma_{k \in M(l)} X_{lk} \qquad (1)$$

applied separately to $X = TT$ (giving the mean travel time $TT_l$) and $X = C$ (giving the mean $CO_2$ $C_l$), where $M(l)$ is the set of vehicles whose report on segment $l$ is still inside the RSU's current averaging window. It is $TT_l$ and $C_l$ (not the raw per-vehicle $TT_{lk}$, $C_{lk}$) that are exchanged between RSUs in Phase 2 and compared in Phase 3.

**Phase 2: I2I, gossip.** RSUs periodically broadcast their accumulated per-segment averages to neighboring RSUs, so that knowledge of a congested or clean segment propagates beyond the RSU that directly observed it.

**Phase 3: V2I / I2V, route request/answer.** A vehicle entering an RSU's range sends a route request; the RSU compares the outgoing segments it knows about using a weighted combination of mean time, mean $CO_2$, and distance, and answers with routing guidance.

The original decision rule is a weighted sum: normalize each candidate segment's time, $CO_2$, and distance, combine them by fixed weights, and select the minimum. The follow-up work [2] replaces this with a majority-voting comparison across the same three features to reduce sensitivity to any single tweaked report; we implement the original weighted-sum rule, discussed further in Sections IV and VIII.

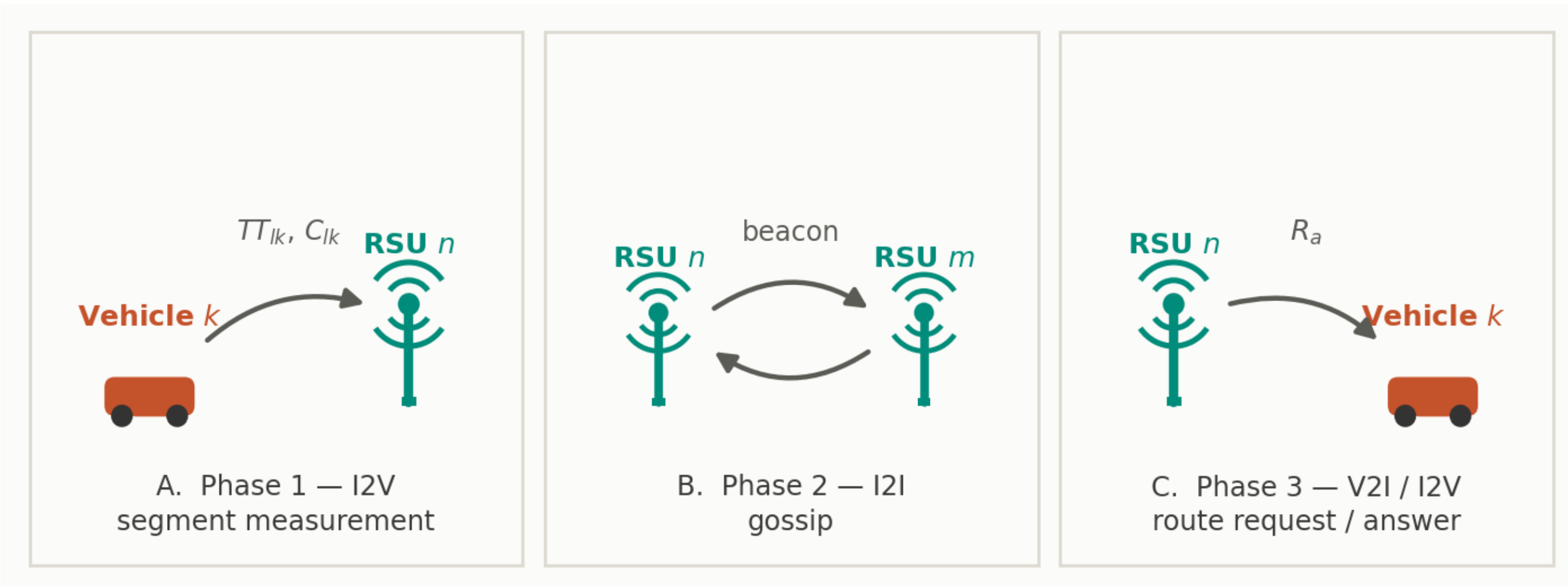

Fig. 2. ErouVe's three communication phases between a vehicle and its serving RSU(s).

## IV. SYSTEM DESIGN

To evaluate ErouVe under controlled, varying road architectures and traffic regimes, we require a simulator that efficiently executes its three-phase protocol. This section describes that evaluation testbed, the two further weaknesses it revealed once the protocol started exchanging messages under load, and the corrective methods we propose for each.

### *A. Platform*

Our evaluation testbed is a Veins subproject (veins_erouve) linking against Veins 5.2 [3], running on OMNeT++ 5.7 with SUMO 1.11 [4] as the road-traffic backend, coupled over TraCI. Vehicles and RSUs run IEEE 802.11p at 6 Mbps and 20 mW, with a 2 s RSU gossip interval.

### *B. Adapting a Paper Protocol to a Real Router*

Veins exposes changeRoute(edgeId, travelTime), which sets a per-vehicle override on one edge's cost and triggers SUMO's own Dijkstra-based rerouting, but it does not expose the road graph's connectivity, so an RSU cannot literally enumerate outgoing segments from a junction the way the original protocol description implies. We route around this by having each RSU answer a request with adapted costs for every segment it has data on; the requesting vehicle applies each cost via changeRoute() and lets SUMO's own router find the minimum-cost path over the remainder of the trip, not just the next hop. This is strictly more general than the original per-junction comparison and required no manual shortest-path implementation of our own.

### *C. Weakness 1: A Cost Metric That Rewards Idling*

A segment's own observed mean travel time $TT_l$ is already a fair time cost, since it reflects both length and any real congestion, so we use it directly rather than normalizing it against other segments, which would unfairly penalize a longer segment regardless of how clean it is. The protocol's $CO_2$ term is a different matter. Taken directly as a per-second rate, as the original description implies, it is diluted by idling time, during which a vehicle stopped at a red light emits almost nothing, so a congested signalized segment can misleadingly appear clean by that measure even though it burns far more fuel to cross the same distance. This is a weakness in the cost metric itself, not merely an implementation detail: any per-second emissions signal rewards standing still.

***We propose, and use throughout, a distance-normalized $CO_2$ cost***: emissions per meter actually driven, min–max normalized against the other segments the RSU currently knows about. Let $len_l$ denote segment $l$'s length in meters, so that $c_l = C_l / len_l$ (Eq. (1)'s $C_l$) is its mean emission rate per meter driven. Writing $\Omega$ for the set of segments the requesting RSU currently holds data for, the normalized cost used in the routing decision is

$$\hat{c}_l = (c_l - \min_{l' \in \Omega} c_{l'}) / (\max_{l' \in \Omega} c_{l'} - \min_{l' \in \Omega} c_{l'}) \qquad (2)$$

where:

- $c_l$ = per-meter $CO_2$ rate of the specific segment the protocol is computing the normalized cost for
- $c_{l'}$ = per-meter $CO_2$ rate of some other segment l' in $\Omega$ (the set of segments the RSU currently has data for)
- l' sweeps over every member of $\Omega$, including l itself, to find the best (min) and worst (max) rate currently known

which maps the best segment the RSU currently knows about to $\hat{c} = 0$ and the worst to $\hat{c} = 1$; $\hat{c}_l$ is the $CO_2$ term

ErouVe's decision rule combines with time and distance. We tried max-only normalization first (dividing $c_l$ by its maximum rather than min–max scaling it), and it failed, because real per-meter $CO_2$ values cluster in a narrow band nowhere near zero, which compresses the useful signal almost flat.

### *D. Weakness 2: Cold-Start Knowledge Asymmetry*

Two further problems became visible only once the protocol was actually exchanging messages under load, and we state both generally, since they apply to any system that lets vehicles act on a crowd-sourced, incrementally-built estimate.

***Weakness.*** An RSU that has learned a clean alternative before it has learned the congested baseline will confidently route vehicles off a good road and onto a bad one it has not measured yet, because an unmeasured segment defaults to its optimistic, uncongested static weight, which looks cheap purely due to lack of data.

***Proposed method: knowledge-gated cost acceptance.*** We never let an RSU adjust an alternative's cost unless it also already knows the cost of the vehicle's own currently-planned next segment, so that an unmeasured, possibly-bad edge is never silently trusted at its default weight.

### *E. Weakness 3: Decision Flapping*

***Weakness.*** Re-applying a fresh cost estimate on every RSU gossip reception (roughly every 2 s) let a later, noisier, lower-sample-count estimate undo an earlier, correct diversion, with whichever estimate happened to arrive last, rather than the best one, determining the vehicle's final route.

***Proposed method: diversion locking.*** We have a vehicle stop reprocessing further gossip for its current segment once it diverts from its default plan: locking in a genuine improvement is more stable than continuously second-guessing it.

Section VI evaluates the protocol with all three proposed methods active; Section VII discusses how completely each closes the weakness it targets, including a residual case that knowledge-gating and diversion-locking do not fully resolve.

## V. EXPERIMENTAL SETUP

Reproducing ErouVe's own evaluation network was not our goal; we needed a network with a genuine, measurable choice in it. Veins' bundled Erlangen example, used in our early testing, turned out to route effectively every vehicle down one shared corridor with no real alternative, so no rerouting decision was ever available to make.

### *A. Network Topology*

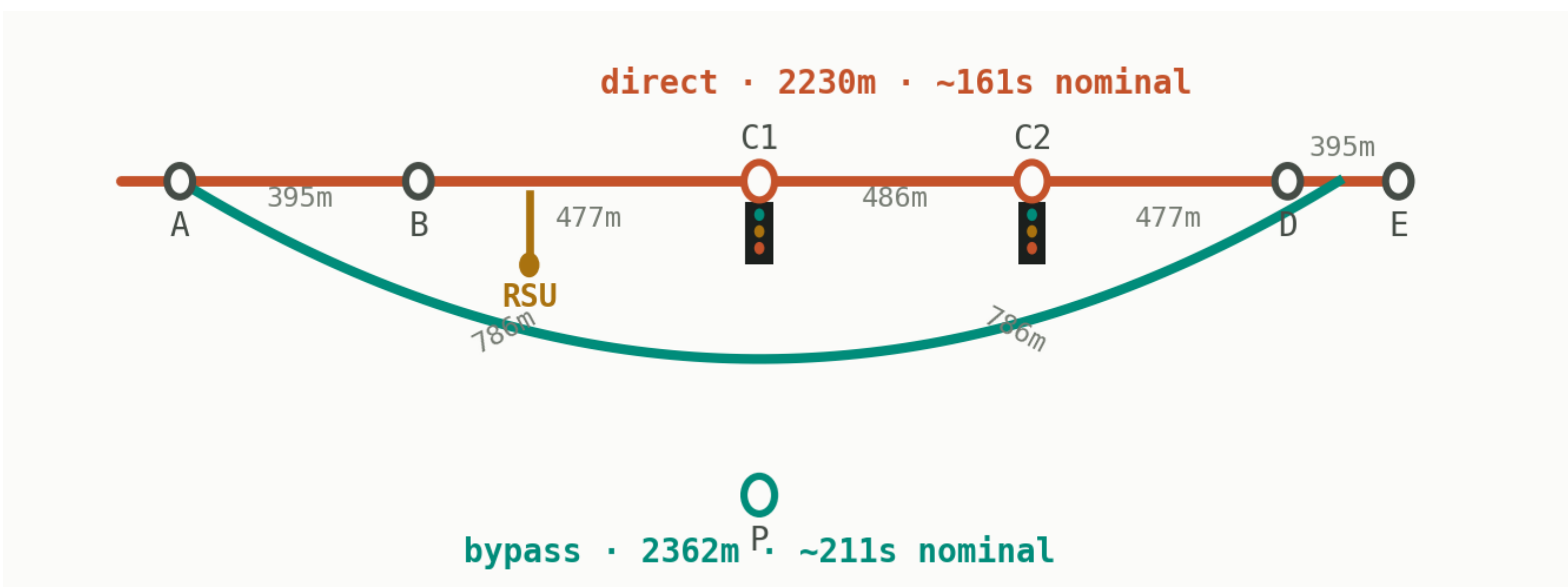

Fig. 3. Network topology. The direct route (A–B–C1–C2–D–E, orange) is nominally faster and is what plain SUMO routing selects by default, but crosses two signalized junctions. The bypass (through P, teal) is 6% longer with a lower speed limit, but signal-free. One RSU sits near the network's center, in radio range of both alternatives at the diverge point.

TABLE I EDGE GEOMETRY

| Segment | Route | Length | Speed limit | Notes |
|---|---|---|---|---|
| AB | shared | 395 m | 50 km/h | entry leg |
| B–C1–C2–D | direct | 477/486/477 m | 50 km/h | 2 signalized jct. |
| DE | shared | 395 m | 50 km/h | exit leg |
| B–P–D | bypass | 786/786 m | 40 km/h | signal-free |
| Direct total | — | 2230 m | — | nominal ~161 s |
| Bypass total | — | 2362 m | — | nominal |
| | | | | ~211 s (+31%) |

Both junctions run a static 90 s program (42 s green, 3 s yellow per direction), giving the through movement roughly 47% effective green share, a one-lane saturation throughput of about 0.24 veh/s (≈880 veh/h) which is rather optimistic. This could also be a parameter that could affect the results in future tests.

### *B. Traffic Demand*

We ran two scenarios over the road network and protocol. Light keeps demand comfortably under the signal's throughput capacity, so delay is limited to each vehicle independently waiting out one red phase. Heavy pushes mainFlow's arrival rate to roughly twice that capacity, so a real, growing queue forms behind C1 instead.

TABLE II VEHICLE FLOWS

| Regime | Flow | Route | Window | Period | Rate | n |
|---|---|---|---|---|---|---|
| Light | explore | bypass | 0–450s | 10 s | 0.10 v/s | 45 |
| Light | main | direct | 200–450s | 5 s | 0.20 v/s | 50 |
| Heavy | explore | bypass | 0–500s | 8 s | 0.125 v/s | 63 |
| Heavy | main | direct | 200–500s | 2 s | 0.50 v/s | 150 |

exploreFlow always starts well before mainFlow: the bypass leg alone takes 70–80 s to traverse, so without a head start the RSU has no real bypass data yet by the time the first mainFlow vehicles reach the diverge point. All vehicles share one type: 5 m length, 2.6/4.5 m/s² accel/decel, 0.5 driver-imperfection sigma, HBEFA3/PC_G_EU4 emission class, i.e., a conventional petrol passenger car, consistent with the fleet-composition point made in Section I.

### *C. Protocol Parameters*

TABLE III ErouVe CONFIGURATION

| Parameter | Value | Applies to |
|---|---|---|
| timeWeight | 0.2 | RSU & vehicle cost fn. |
| co2Weight | 0.8 | RSU & vehicle cost fn. |
| gossipInterval | 2 s | RSU I2I broadcast |
| requestInterval | 5 s | vehicle V2I throttle |
| RSU count/placement | 1 | network center, near C1 |

## VI. RESULTS

### *A. Isolated Route Comparison (No Rerouting)*

Before evaluating the protocol, we confirmed the premise it depends on: running each flow on its assigned route with no rerouting at all, direct SUMO accounting shows the bypass is genuinely cleaner per trip, and that the gap is traffic-dependent.

TABLE IV ORACLE ROUTE ASSIGNMENT, MEAN $CO_2$ PER VEHICLE

| Regime | Route | n | Mean $CO_2$ | Gap |
|---|---|---|---|---|
| Light | direct | 50 | 475,143 mg | −8.9% |
| Light | bypass | 45 | 433,029 mg | — |
| Heavy | direct | 150 | 605,075 mg | −20.5% |
| Heavy | bypass | 63 | 480,890 mg | — |

The bypass's advantage more than doubles under heavy traffic (8.9% → 20.5%), not because the bypass itself changes, but because the direct route's real cost balloons under queueing (maximum observed queue: 29 vehicles; mean direct travel time rose from a nominal 161 s to an observed 339 s).

### *B. Protocol-Driven Results*

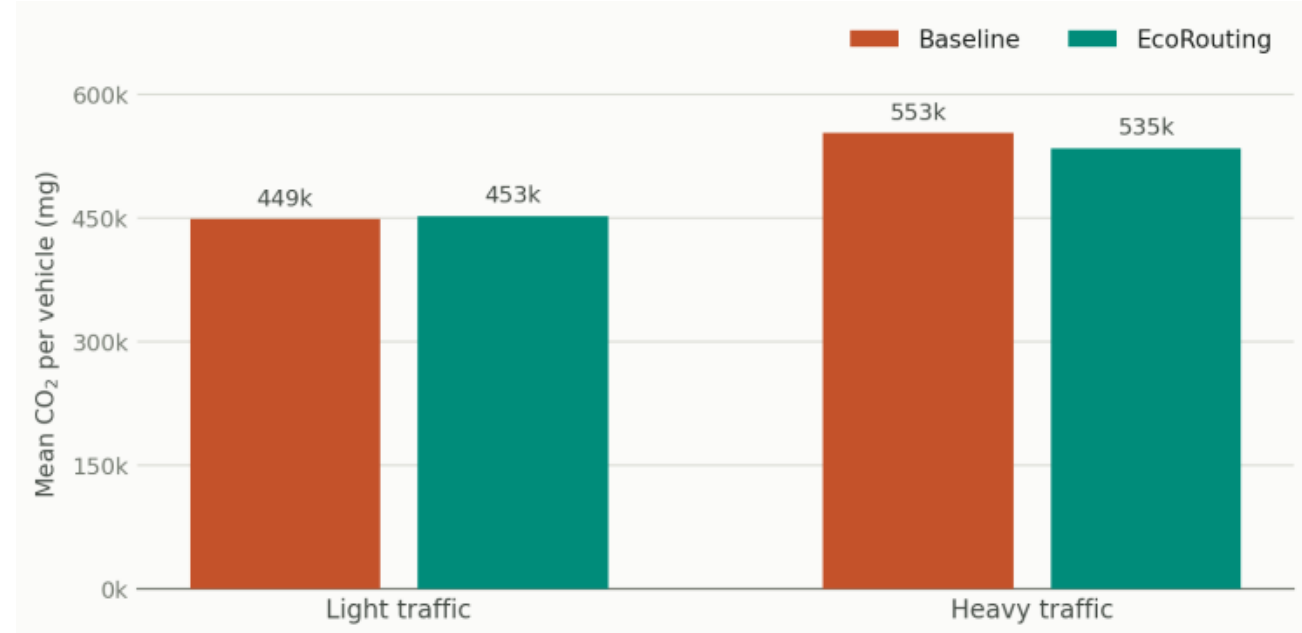

Fig. 4. Fleet mean $CO_2$ per vehicle, Baseline vs. EcoRouting. Under light traffic the two are statistically indistinguishable. Under heavy traffic, EcoRouting reduces the fleet mean by 3.28%.

TABLE V FLEET-WIDE TOTALS

| Regime | Config | n | Total $CO_2$ | Δ |
|---|---|---|---|---|
| Light | Baseline | 95 | 42,657,878 mg | +0.86% |
| Light | EcoRouting | 95 | 43,024,281 mg | — |
| Heavy | Baseline | 213 | 117,854,424 mg | −3.28% |
| Heavy | EcoRouting | 213 | 113,987,890 mg | — |

Splitting the heavy-traffic result by each vehicle's originally-assigned route shows where the reduction comes from, and where the protocol still gives some of it back.

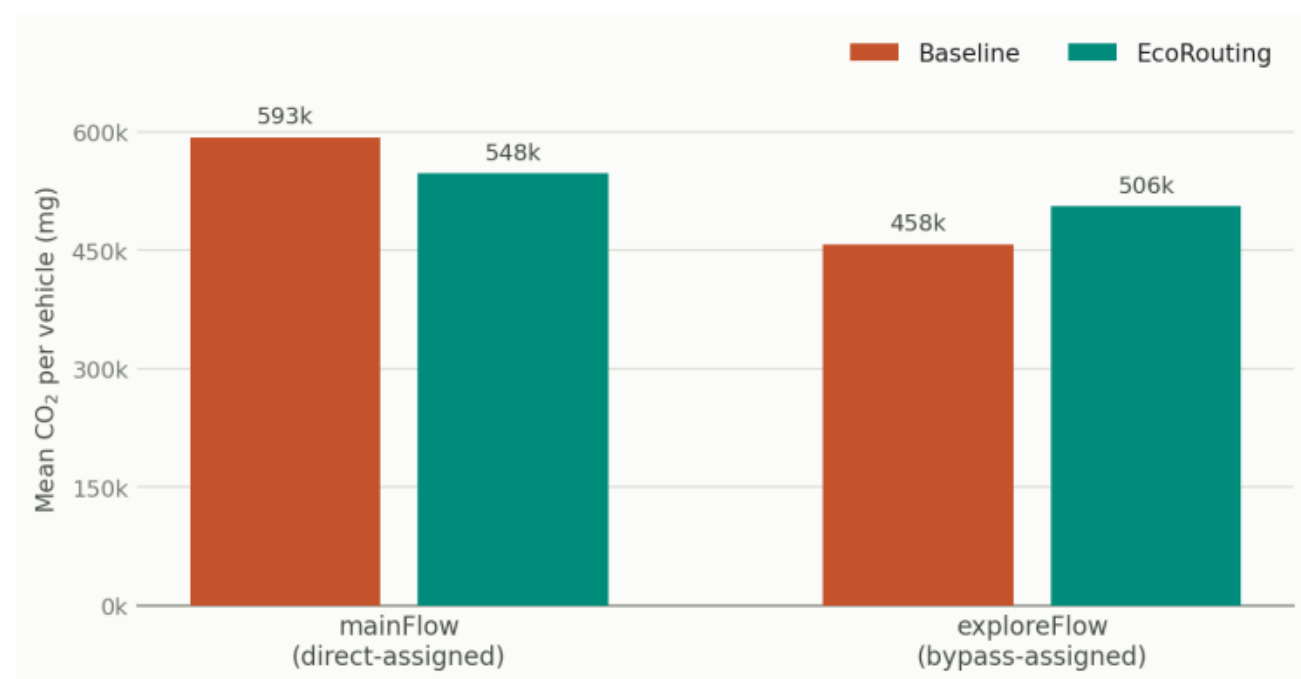

Fig. 5. Heavy-traffic mean $CO_2$, split by each vehicle's original route assignment. mainFlow vehicles (92 of 150 diverted) save 7.7% on average. exploreFlow vehicles (bypass-assigned, 29 of 63 rerouted) regress 10.4%, discussed in Section VII.

TABLE VI HEAVY TRAFFIC, BY ORIGINAL ASSIGNMENT

| Group | n | Rerouted | Baseline | Δ |
|---|---|---|---|---|
| mainFlow (direct) | 150 | 92 (61%) | 593,447 mg | −7.7% |
| exploreFlow (bypass) | 63 | 29 (46%) | 457,737 mg | +10.4% |

## VII. DISCUSSION

***Congestion vs distance.*** The bypass is 6% longer in every run we ran; that never changes. What changes between the light and heavy regimes is whether the direct route's real cost has diverged far enough from its nominal cost to be worth escaping. Under light traffic, staying on the nominally shorter road costs almost nothing extra, so there is very little for a routing correction to recover, and the fleet-wide result is indistinguishable from noise. Under heavy traffic, the queue-induced penalty on the direct route (161 s nominal → 339 s observed) is large enough that redirecting even a majority of one flow produces a measurable fleet-wide reduction. This is a specific, testable version of a general point: an eco-routing system's ceiling is set by how bad the status quo currently is, not by how clean the alternative is in isolation.

***ExploreFlow regression.*** Diversion locking (Section IV-E) fixed vehicles flapping back and forth mid-decision, and knowledge-gated cost acceptance (Section IV-D) fixed the worst cold-start cases. Both proposed methods were necessary but not sufficient: 29 of 63 bypass-assigned vehicles still end up rerouted, and on balance that group's mean $CO_2$ rises rather than falls. The most likely remaining cause is the same asymmetry in a subtler form: a bypass-bound vehicle can still receive a gossip update in which the direct road's partial known cost (say, just the first signalized segment) briefly looks cheaper than the bypass's own real cost, before the rest of the direct road's congestion has been reported back. Tightening this would mean requiring more of the competing route to be known, not just the vehicle's own next hop, which we regard as a natural next iteration rather than a fundamental limitation.

***Fleet composition, revisited.*** The emission model we use throughout (HBEFA3/PC_G_EU4) represents a petrol passenger car, deliberately, because, as we noted in Section I, that is still the overwhelming majority of what is actually moving through cities today. A 3.28% fleet-wide reduction sounds modest until it is weighed against how much of the current urban fleet it is addressable to: not a shrinking legacy segment, but upwards of 95% of vehicles on the road [6].

## VIII. CONCLUSION AND FUTURE WORK

In this paper we evaluate ErouVe under road architectures and traffic scenarios which gave us a clear picture of the protocol's strengths and weaknesses. Its central strength is substantial: under traffic dense enough to saturate a signalized route, ErouVe recovers most of the route $CO_2$ gap, cutting the fleet-wide mean by 3.28% and diverted vehicles' own emissions by 7.7%, with only a one-line change to how routing decisions are applied. Its weaknesses were revealed once the protocol was exercised under diverse load scenarios: a per-second $CO_2$ metric that rewards idling, a cold-start knowledge asymmetry, and decision flapping under noisy crowd-sourced samples.

We proposed one corrective method for each problem identified; a distance-normalized cost function, knowledge-gated cost acceptance, and diversion locking. We prove that all three measurably restored the protocol effectiveness.

We consider the following future directions and research paths:

1. closing the exploreFlow regression identified in Section VII by requiring fuller knowledge of the competing route before diverting.
2. testing the majority-voting decision rule from the follow-up [2] against the same scenarios.
3. extending beyond a single two-route topology to a network with multiple simultaneous diverge points and multi-hop I2I communications, to test whether our proposed methods generalize to more complex road architectures;
4. re-running this same light/heavy comparison on a real, OpenStreetMap-derived road network.

## REFERENCES

[1] L. A. Maglaras, P. Basaras, and D. Katsaros, “Exploiting Vehicular Communications for Reducing $CO_2$ Emissions in Urban Environments,” in Proc. 2013 Int. Conf. Connected Vehicles and Expo (ICCVE), IEEE, 2013, pp. 32–37.

[2] P. Basaras, L. Maglaras, D. Katsaros, and H. Janicke, “A Robust Eco-Routing Protocol Against Malicious Data in Vehicular Networks,” in Proc. 8th IFIP Wireless and Mobile Networking Conf. (WMNC), 2015, pp. 184–191.

[3] C. Sommer, R. German, and F. Dressler, “Bidirectionally Coupled Network and Road Traffic Simulation for Improved IVC Analysis,” IEEE Trans. Mobile Computing, vol. 10, no. 1, pp. 3–15, Jan. 2011.

[4] P. A. Lopez, M. Behrisch, L. Bieker-Walz, J. Erdmann, Y.-P. Flötteröd, R. Hilbrich, L. Lücken, J. Rummel, P. Wagner, and E. Wießner, “Microscopic Traffic Simulation using SUMO,” in Proc. 21st IEEE Int. Conf. Intelligent Transportation Systems (ITSC), 2018, pp. 2575–2582.

[5] International Energy Agency, Global EV Outlook 2025, IEA, Paris.

[6] International Energy Agency, Global EV Outlook 2026, IEA, Paris.

[7] A. F. Santamaria, C. Sottile, F. De Rango, and S. Marano, “Safety Enhancement and Carbon Dioxide ($CO_2$) Reduction in VANETs,” Mobile Networks and Applications, vol. 20, no. 2, pp. 220–238, 2015.

[8] C. T. Barba, M. A. Mateos, P. R. Soto, A. M. Mezher, and M. A. Igartua, “Smart City for VANETs Using Warning Messages, Traffic Statistics and Intelligent Traffic Lights,” in Proc. IEEE Intelligent Vehicles Symp. (IV), 2012, pp. 902–907.

[9] S. Tsugawa and S. Kato, “Energy ITS: Another Application of Vehicular Communications,” IEEE Communications Magazine, vol. 48, no. 11, pp. 120–126, 2010.

[10] A. M. de Souza, A. Boukerche, G. Maia, R. I. Meneguette, A. A. Loureiro, and L. A. Villas, “Decreasing Greenhouse Emissions Through an Intelligent Traffic Information System Based on Inter-vehicle Communication,” in Proc. 12th ACM Int. Symp. Mobility Management and Wireless Access (MobiWac), 2014, pp. 91–98.

[11] M. Milojevic and V. Rakocevic, “Distributed Road Traffic Congestion Quantification Using Cooperative VANETs,” in Proc. 13th IFIP/IEEE Annu. Mediterranean Ad Hoc Networking Workshop (Med-Hoc-Net), 2014, pp. 203–210.

[12] L. A. Maglaras, “A Novel Distributed Intrusion Detection System for Vehicular Ad Hoc Networks,” Int. J. Advanced Computer Science and Applications, vol. 6, no. 4, pp. 101–106, 2015.

[13] F. M. Talaat, W. M. Shaban, H. ZainEldin, et al., “Integrating V2X Solutions in Intelligent Green Cities: An AI-Driven Point Exchange System Approach,” Neural Computing and Applications, vol. 37, pp. 15311–15343, 2025.

[14] H. Alabdouli, M. S. Hassan, and A. Abdelfatah, “Enhancing Route Guidance with Integrated V2X Communication and Transportation Systems: A Review,” Smart Cities, vol. 8, no. 1, p. 24, 2025.

[15] X. Wang, H. Mu, Y. Fan, et al., “Road and Speed Restriction Strategies for Alleviating Tunnel Congestion Based on SUMO Traffic Simulation,” Int. J. Intelligent Transportation Systems Research, 2025.

[16] T. Liang, T. Bai, and A. A. Malikopoulos, “A Coordinated Routing Approach for Enhancing Bus Timeliness and Travel Efficiency in Mixed-Traffic Environment,” arXiv:2505.01566, 2025.

[17] E. Mármol Campos, J. L. Hernandez-Ramos, and A. González Vidal, “Misbehavior Detection in Intelligent Transportation Systems Based on Federated Learning,” Internet of Things, vol. 25, art. 101127, 2024.

[18] F. Mansouri, M. Tarhouni, and B. Alaya, “A Distributed Intrusion Detection Framework for Vehicular Ad Hoc Networks via Federated Learning and Blockchain,” Ad Hoc Networks, vol. 167, art. 103677, 2025.